\documentclass[11pt]{article}
\usepackage{epsfig}

\newcommand{\beq}{\begin{equation}}
\newcommand{\eeq}{\end{equation}}
\newcommand{\beqa}{\begin{eqnarray}}
\newcommand{\eeqa}{\end{eqnarray}}

\newcommand{\NPB}[1]{{\it Nucl. Phys.}\ {\bf B{#1}}}
\newcommand{\PLB}[1]{{\it Phys. Lett.}\ {\bf B{#1}}}
\newcommand{\PRD}[1]{{\it Phys. Rev.}\ {\bf D{#1}}}
\newcommand{\PRL}[1]{{\it Phys. Rev. Lett.}\ {\bf #1}}
\newcommand{\NCA}[1]{{\it Nuovo Cim.}\ {\bf {#1}A}}

\newcommand{\laem}{\stackrel{<}{\sim}}
\newcommand{\gaem}{\stackrel{>}{\sim}}

\newcommand{\tr}{{\rm Tr}}

\begin{document}

\begin{titlepage}
\def\thepage {}        

\title{Flavor Physics and Fine-Tuning in Theory Space}

\author{
R. Sekhar Chivukula$^1$, Nick Evans$^3$, 
and Elizabeth H. Simmons$^{1,2}$\thanks{e-mail addresses
  sekhar@bu.edu, evans@phys.soton.ac.uk, simmons@bu.edu}\\ \\
$^1$Department of Physics, Boston University, \\
590 Commonwealth Ave., Boston MA  02215 \\ \\
$^2$Department of Physics, Harvard University, \\
Cambridge, MA, 02138 \\ \\
$^3$Dept. of Physics, University of Southampton, \\                  
Highfield, Southampton, SO17 1BJ, U.K.}

\date{April 26, 2002}

\maketitle

\bigskip
\begin{picture}(0,0)(0,0)
\put(355,295){BUHEP-02-17}
\put(355,280){HUTP-02/A010}
\put(355,265){hep-ph/0204193}
\end{picture}
\vspace{24pt}

\begin{abstract}

Recently a new class of composite Higgs models have been developed
which give rise to naturally light Higgs bosons without
supersymmetry. Based on the chiral symmetries of ``theory space,''
involving replicated gauge groups and appropriate gauge symmetry
breaking patterns, these models allow the scale of the underlying
strong dynamics giving rise to the composite particles to be as large
as of order 10 TeV, without any fine tuning to prevent large
corrections to Higgs boson mass(es) of order 100 GeV.  In this note we
show that the size of flavor violating interactions arising
generically from underlying flavor dynamics constrain the scale of the
Higgs boson compositeness to be greater than of order 75 TeV, implying
that significant fine-tuning is required. Without fine-tuning, the
low-energy structure of the composite Higgs model alone is not
sufficient to eliminate potential problems with flavor-changing
neutral currents or excessive CP violation; solving those problems
requires additional information or assumptions about the symmetries of
the underlying flavor or strong dynamics. We also consider the weaker,
but more model-independent, bounds which arise from limits on weak
isospin violation.

\pagestyle{empty}
\end{abstract}
\end{titlepage}

\setcounter{section}{0}
\setcounter{equation}{0}
\setcounter{footnote}{0}

\bigskip

\section{Introduction}

Recently a new class \cite{Arkani-Hamed:2001nc, Arkani-Hamed:2002pa}
of composite Higgs models \cite{chiggs} has been developed which give
rise to naturally light Higgs bosons without supersymmetry. Inspired
by discretized versions of higher-dimensional gauge theory
\cite{Arkani-Hamed:2001ca, Cheng:2001nh}, these models are based on
the chiral symmetries of ``theory space''\cite{Arkani-Hamed:2001ca}
. The models involve replicated gauge groups and corresponding gauge
symmetry breaking patterns.   They allow the scale ($\Lambda$) of the
underlying strong dynamics giving rise to the composite particles to
be as large as 10 TeV, without causing large corrections to the Higgs
boson mass(es) of order 100 GeV.

Various possibilities exist for the underlying physics (the
``high-energy completion'') which gives rise to the chiral-symmetry
breaking pattern required, and produces the ``pion'' which becomes the
composite Higgs.  However, regardless of the precise nature of the
underlying strongly-interacting physics, there must be flavor dynamics
at a scale of order $\Lambda$ or greater that gives rise to the
different Yukawa couplings of the Higgs boson to ordinary fermions.
As in extended technicolor theories \cite{etci,etcii}, if this flavor
dynamics arises from gauge-interactions it will generically cause
flavor-changing neutral currents \cite{etcii}.  

In this note we review and update the lower bound on $\Lambda$ arising
from the experimental constraints on extra contributions to the
neutral meson mass differences \cite{Chivukula:1997iw}.  We find that
in composite Higgs models the size of flavor-violating interactions
arising from the high-energy theory constrain the scale $\Lambda$ to
be greater than of order 75 TeV.  We then consider the ``theory
space'' models, argue why this flavor bound applies to such models,
and review the upper limit on $\Lambda$ of order 10 TeV necessary to
avoid fine-tuning \cite{Arkani-Hamed:2001nc,
Arkani-Hamed:2002pa}. Raising the scale $\Lambda$ to 75 TeV to be
consistent with the flavor bounds mentioned above, then, necessitates
fine tuning of order 2\%. We compare these bounds to those arising
from limits on the amount of CP violation and isospin violation in the
composite Higgs theory.

The implication of our findings is that the low-energy structure of
the composite Higgs model alone is not sufficient to eliminate
potential problems with flavor-changing neutral current or excessive
CP violation; solving those problems requires additional information
or assumptions about the symmetries of the underlying strong
dynamics\footnote{See also \protect\cite{Lane:2002pe}, which
emphasizes that the properties of the underlying strong-dynamics may
affect the details of the low-energy phenomenology.}.

\section{Flavor and Composite Higgs Bosons\protect\footnote{This section  
reviews and updates material from \protect\cite{Chivukula:1997iw}.}}
\label{sec:flavor}

We begin by considering what the observed masses of the ordinary
fermions imply about the underlying flavor physics. Providing the different
masses of the fermions requires flavor physics (analogous to
extended-technicolor interactions (ETC) \cite{etci,etcii}) which couples
the left-handed quark doublets $\psi_L$ and right-handed singlets $q_R$
to the strongly-interacting constituents of the composite Higgs doublet.
At low energies, these interactions produce the quark Yukawa couplings.

To estimate the sizes of various effects of the underlying physics, we
rely on dimensional analysis \cite{ndaref}. As noted by Georgi
\cite{generalized}, a theory with light scalar particles belonging to
a single symmetry-group representation depends on two parameters:
$\Lambda$, the scale of the underlying physics, and $f$ (the analog of
$f_\pi$ in QCD), which measures the amplitude for producing the scalar
particles from the vacuum. Our estimates of the sizes of the
low-energy effects of the underlying physics will depend on the ratio
$\kappa \equiv \Lambda / f$, which determines the sizes of coupling
constants in the low-energy theory. Naive dimensional analysis corresponds
to $\kappa = 4\pi$ \cite{ndaref}.

Assuming that these new flavor interactions are gauge interactions
with gauge coupling $g$ and gauge boson mass $M$, dimensional analysis
\cite{ndaref} allows us to estimate that the size of the resulting
Yukawa coupling is \cite{chiggs} of order
$(g^2/M^2)(\Lambda^2/\kappa)$, i.e.
\beq {\lower35pt\hbox{\epsfysize=1.0 truein
    \epsfbox{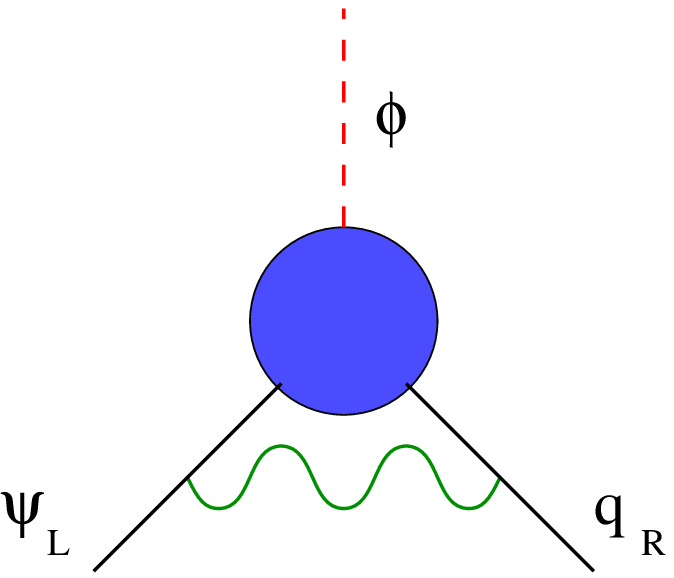}}} \Rightarrow {g^2 \over M^2} {\Lambda^2 \over
  \kappa}\bar{q}_R \phi \psi_L ~.
\label{eq:quarkpreon}
\eeq
In order to give rise to a quark mass $m_q$, the 
Yukawa coupling must be equal to
\beq
{\sqrt{2} m_q \over v}
\eeq
where $v\approx 246$ GeV. This implies
\beq
\Lambda \approx {M \over g} \sqrt{\sqrt{2} \kappa {m_q \over v}}~.
\label{eq:yukawa}
\eeq 
Thus, if we set a lower limit on $M/g$ from low-energy flavor physics,
eqn.(\ref{eq:yukawa}) will give a lower bound on $\Lambda$.

The high-energy flavor physics responsible for the generation of the
Yukawa couplings {\it must} distinguish between different flavors so
as to give rise to the different masses of the corresponding fermions.
In addition, the flavor physics will give rise to flavor-specific
couplings among ordinary fermions \cite{etci,etcii}.  These will generically
give rise to flavor-changing neutral currents (as previously noted in
\cite{etcii} for the case of ETC theories) that affect Kaon, $D$-meson,
and $B$-meson physics. 

Consider the interactions responsible for the $c$-quark mass. Through
Cabibbo mixing, these interactions must couple to the $u$-quark as
well.  Neglecting mixing with the top-quark, this will generally give
rise to the interactions
\beqa
{\cal L}_{eff} & = & - \, (\cos \theta_L^c \sin \theta_L^c)^2 
\frac{g^{2}}{M^{2}}
( \overline c_L  \gamma^{\mu} u_L )(\overline c_L  \gamma_{\mu} u_L)
\nonumber \\ [2mm]
& & -  \, (\cos \theta_R^c \sin \theta_R^c)^2 
\frac{g^{2}}{M^{2}}
( \overline c_R \gamma^{\mu} u_R)(\overline c_R \gamma_{\mu} u_R)
\nonumber \\ [2mm]
& & - \,
2 \cos \theta_L^c \sin \theta_L^c \cos \theta_R^c \sin \theta_R^c
\frac{g^{2}}{M^{2}}
( \overline c_L  \gamma^{\mu} u_L )(\overline c_R \gamma_{\mu} u_R)~,
\label{ops1}
\eeqa
where the coupling $g$ and mass $M$ are of the same order as those in
the interactions which ultimately give rise to the $c$-quark Yukawa
coupling in eqn.~(\ref{eq:quarkpreon}), and the angles $\theta^c_L$
and $\theta^c_R$ represent the relation between the gauge eigenstates
and the mass eigenstates.  The operators in eqn.~(\ref{ops1}) will
clearly affect neutral D-meson physics.  Similarly, the interactions
responsible for other quarks' masses will give rise to operators that
contribute to mixing and decays of the corresponding mesons.

The color-singlet products of currents in eqn.~(\ref{ops1}) will
contribute directly to $D$-meson mixing.  In the vacuum-insertion
approximation, the purely left-handed or right-handed current-current
operators yield
\beq
\left(\frac{M}{g}\right)_{ \! {\rm LL,RR}} 
\gaem 
f_D\left( \frac{2  m_D B_D}{3 \Delta m_D }\right)^{\! 1/2}
\cos \theta_{L,R}^c \sin \theta_{L,R}^c \approx 225 \, {\rm TeV} ~,
\eeq
where we have used the limit on the neutral $D$-meson mass difference,
$\Delta m_D \laem  4.6 \times  10^{-11}$ MeV \cite{PDG},
and $f_D \sqrt{B_D} = 0.2$ GeV \cite{FDREF}, $\theta_{L,R}^c \approx \theta_C$.
The bound on the scale of the underlying strongly-interacting
dynamics follows from eqn.~(\ref{eq:yukawa}): 
\beq
\Lambda \gaem 21 \, {\rm  TeV}
\sqrt{\kappa\left({m_c\over 1.5\, {\rm GeV}}\right)}~,
\label{eq:Dbound}
\eeq
so that $\Lambda \gaem 75$ TeV for $\kappa \approx 4\pi$.

The $\Delta C = 2$, LR product of color-singlet currents gives a
weaker bound than eqn.~(\ref{eq:Dbound}), but the LR product of
color-octet currents,
\beq
{\cal L}_{eff} = - \,
2 \cos \theta_L^c \sin \theta_L^c \cos \theta_R^c \sin \theta_R^c
\frac{g^2}{M^2}
( \overline c_L \gamma^{\mu} T^a u_L)
(\overline c_R \gamma_{\mu} T^a u_R) ~,
\label{ops2}
\eeq
where $T^a$ are the generators of $SU(3)_C$, gives a stronger bound:
\beqa
\left(\frac{M}{g}\right)_{ \! {\rm LR}} & \gaem &
\frac{4 f_D}{3(m_c + m_u)} \left( \frac{m_D^3 B_D^\prime}{\Delta
  m_D}\right)^{\! 1/2}
(2 \cos \theta_L^c \sin \theta_L^c \cos \theta_R^c \sin \theta_R^c)^{\!
  1/2}
\\ [2mm]
& \approx & 590 \, {\rm TeV} \left({1.5\, {\rm GeV}\over m_c}\right)~,
\eeqa
corresponding to 
\beq
\Lambda \gaem 53 \, {\rm  TeV}
\sqrt{\kappa\left({1.5\, {\rm GeV}\over m_c}\right)}~.
\label{eq:DDbound}
\eeq

There are also contributions to $K$-meson mixing from the
color-singlet and color-octet products of currents analogous to those in
eqns.~(\ref{ops1}) and (\ref{ops2}).  The lower bound on
$\Lambda$ derived from the measured value of the $K_L K_S$ mass
difference \cite{Chivukula:1997iw}
\beq
\Lambda \gaem 6.8 \, {\rm TeV}
\sqrt{\kappa\left({m_s\over 200\, {\rm MeV}}\right)}~.
\label{eq:fcncbound}
\eeq
is weaker than (\ref{eq:Dbound}) because the $s$-quark is
lighter than the $c$-quark, while the $d-s$ and $u-c$ mixings are
expected to be of comparable size
\cite{Chivukula:1997iw}.  However, in the absence of additional
superweak interactions to give rise to CP-violation in $K$-mixing
($\varepsilon$), the flavor interactions responsible for the $s$-quark
Yukawa couplings must violate CP at some level. In this case the the
bounds on the scale $\Lambda$ are much stronger.  Recalling that
\beq
 {\rm Re}\, \varepsilon \approx  { {\rm Im M_{12}} \over {2\, \Delta M}} 
 \laem 1.65\,\times\,10^{-3}\, ,
\eeq
and assuming that there are phases of order 1 in the $\Delta S=2$
operators analogous to those shown in eqn. (\ref{ops1}), we find the
bound
\beq
\Lambda \gaem 120\, {\rm TeV} \sqrt{\kappa \left({m_s \over 200\, 
{\rm MeV}}\right)}~.
\label{eq:Ebound}
\eeq

\section{Composite Higgs Bosons from Theory Space}

\begin{figure}[bt]
\begin{center}
\includegraphics[width=10cm]{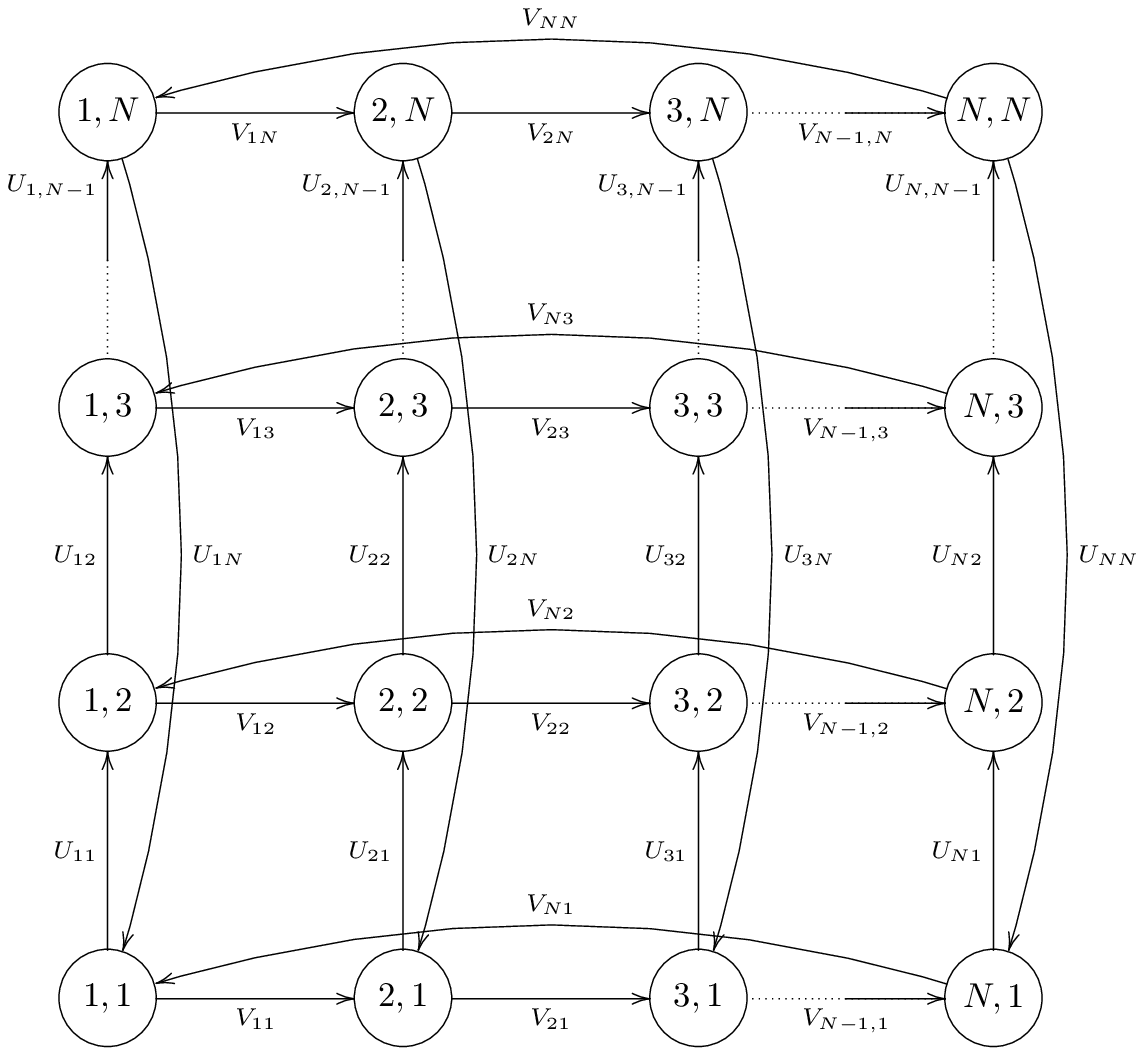} %
\end{center}
\caption{A composite Higgs model based
on an $N \times N$ toroidal lattice ``theory space.'' $SU(3)$ gauge
groups live at every site except $(1,1)$, while the links represent
non-linear sigma fields transforming as $(3,\bar{3})$'s under the
adjacent gauge symmetries.  Only an $SU(2) \times U(1)$ subgroup of an
$SU(3)$ global symmetry group is gauged at site $(1,1)$. As described
in the text, $N^2-1$ sets of Goldstone bosons are eaten, $N^2-1$ get
mass from ``plaquette operators'' which explicitly break the chiral
symmetries, and two sets remain in the very low-energy
theory.  This illustration comes from \protect\cite{Lane:2002pe}.}
\label{fig:moose}
\end{figure}

A set of ``theory space'' composite Higgs models 
\cite{Arkani-Hamed:2001nc, Arkani-Hamed:2002pa} is illustrated in
Figure \ref{fig:moose}, using ``moose'' or ``quiver'' notation
\cite{moose}. In this diagram, each site except $(1,1)$ represents a
gauged $SU(3)$ group, while the links represent non-linear sigma
fields transforming as $(3,\bar{3})$'s under the adjacent groups:
\beq
U_{ij} \to W_{ij}U_{ij}W^\dagger_{i~j+1}~,
\hskip+2cm
V_{ij} \to W_{ij} V_{ij} W^\dagger_{i+1~j}~.
\eeq
The ``toroidal'' geometry of theory space implies that the indices
$i,j$ are periodic mod $N$.  At the site $(1,1)$, only the
$SU(2)\times U(1)$ subgroup of an $SU(3)$ global symmetry is
gauged. The kinetic energy terms in the Lagrangian then read
\beq
{\cal L}_{kin} = - \sum_{ij} {1\over 2 g_{ij}^2} \tr F^2_{ij}
+ {f^2\over 4} \sum_{ij} \tr |D^\mu U_{ij}|^2 +
 {f^2\over 4} \sum_{ij} \tr |D^\mu V_{ij}|^2~,
\label{kinetic}
\eeq
where $g_{ij}$ are the gauge couplings and $f$ is the ``pion-decay
constant'' of the chiral symmetry breaking dynamics.  For simplicity,
in what follows we will assume that the gauge couplings $g_{ij} = g$
are the same for every site except for $(1,1)$. The rules of naive
dimensional analysis \cite{ndaref} then imply that the scale $\Lambda$ of
the underlying high-energy dynamics which gives rise to this theory is
bounded by of order $4\pi f$.

The $2N^2$ Goldstone bosons of the chiral symmetry breaking dynamics
are incorporated into the sigma-model fields
\beq
U_{ij} = \exp{2i\pi_{u,ij}/f}~,
\hskip+2cm
V_{ij} = \exp{2i\pi_{v,ij}/f}~.
\eeq
The gauge symmetry breaking pattern implied is $SU(3)^{N^2-1} \times
SU(2) \times U(1) \to SU(2)\times U(1)$, resulting in $N^2-1$ sets of
``eaten'' Goldstone bosons. The remaining $N^2+1$ sets of Goldstone
bosons in the physical spectrum interact via the gauge interactions,
which explicitly violate the chiral symmetries. However,
because of the ``topology'' of theory space, the lowest-order
interaction in the effective theory which breaks the chiral-symmetries
in the same way as the gauge interactions only occurs at high order 
\cite{Arkani-Hamed:2001nc}. Therefore, the leading 
contribution to the masses of these remaining scalars from the
low-energy gauge interactions is {\it finite}, and arises at ${\cal
O}(g^4)$ from the Coleman-Weinberg potential \cite{Coleman:jx} .

An important ingredient in these models is a set of nonderivative
chiral-symmetry breaking operators of the form of ``plaquette''
interactions\footnote{Because of the reduced symmetry at site $(1,1)$,
additional operators are present there which play an important role in
the detailed phenomenology of the composite scalar
particles.\cite{Arkani-Hamed:2001nc, Arkani-Hamed:2002pa}}
\beq
{\cal L}_{pl} = \lambda f^4 \sum_{ij} \tr\left(U_{ij} V_{i~j+1}
U^\dagger_{i+1~j} V^\dagger_{ij}\right) + h.c.~,
\eeq
where (again, for simplicity) we have assumed that the dimensionless
coupling constants $\lambda$ are the same for every
plaquette. Expanding these operators in terms of the Goldstone bosons
fields, we find
\beq
{\cal L}_{pl} = -4 \lambda f^2 \sum_{ij} \tr(\pi_{u,ij}+\pi_{v,i~j+1}
-\pi_{u,i+1~j}-\pi_{v,ij})^2 + {\cal O}(\pi^4) +\ldots
\eeq
These operators have the extraordinary feature that they give rise to
masses to $N^2-1$ of the remaining scalars, but leave massless the two
combinations
\beq
\pi_{u,ij} \equiv {U \over N} \hskip+2cm \pi_{v,ij}\equiv {V \over N}~
\label{higgsn}
\eeq
which are uniform in either the ``u'' or ``v'' directions.  The
factors of $N$ arise so as to normalize the $U$ and $V$ fields
correctly. Both
the $U$ and $V$ fields contain $SU(2)\times U(1)$ doublet scalars
$\phi_u$ and $\phi_v$ with the quantum numbers of the Higgs boson. The
theory gives rise to two light (so far in this discussion, massless)
composite Higgs bosons with nonderivative interaction of the form
\cite{Arkani-Hamed:2001nc, Arkani-Hamed:2002pa}
\beq
{\cal L}_{pl} \supset {4\lambda\over N^2} 
\tr(\phi_u \phi^\dagger_u - \phi_v \phi^\dagger_v)^2 + 
{4\lambda\over N^2} (\phi^\dagger _u \phi_u - \phi^\dagger_v \phi_v)^2
~.
\label{twohiggs}
\eeq

Additionally, a negative mass-squared for one or both Higgs bosons may be
introduced either through a symmetry-breaking plaquette operator at
the site $(1,1)$ \cite{Arkani-Hamed:2001nc} or through the effect of
coupling the Higgs bosons to the top-quark \cite{Arkani-Hamed:2002pa}. 
In either case, the resulting mass-squared of the Higgs is 
of order
\beq
|m_h|^2 \simeq {\lambda v^2 \over N^2}~.
\label{higgsmsq}
\eeq

The left- and right-handed quarks and leptons transform under the
$SU(2)\times U(1)$ gauge interactions at the site $(1,1)$
\cite{Arkani-Hamed:2001nc, Arkani-Hamed:2002pa}. For the light
fermions, Yukawa couplings between the fermions and the composite
Higgs bosons are introduced. Such interactions violate the chiral
symmetries protecting the Higgs bosons masses, but the size of the
resulting corrections is small since $m_q
\ll v$. This choice preserves a $(U(2))^5$ flavor symmetry, broken
only by the Yukawa couplings to the composite Higgs, suppressing
flavor-changing neutral currents from the $SU(2)\times U(1)$ and
$SU(3)^{N^2-1}$ gauge bosons.  Because the light quarks obtain mass
from Yukawa couplings to the composite scalars, the bounds on the
compositeness scale derived in Section \ref{sec:flavor} apply to this
model. As noted above, however, the light Higgs boson is
``delocalized'' in theory space, eqn. (\ref{higgsn}), and therefore
has only an amplitude of order $1/N$ of being at site
$(1,1)$. Consequently, we would say that $\Lambda$ must satisfy 
\beq
\Lambda \gaem 21 \, {\rm TeV} \sqrt{\kappa N\,\left({m_c\over 1.5\,
{\rm GeV}}\right)}~,
\label{eq:Dboundn}
\eeq
and be at least of order $\sqrt{N}\cdot 75$ TeV for $\kappa =4\pi$.

The top-quark presents a more difficult problem. In this case, no {\it
direct} Yukawa coupling is introduced \cite{Arkani-Hamed:2001nc}.
Instead, the top-quark is ``spread out'' in theory space: a family of
massive $SU(3)$ vector fermions on the sites $(1,n_v)$ and $(n_u,1)$
is added (here $1\le n_{u,v} \le N$), along with local interactions
between the the vector fermions at adjacent sites and the
gauge-eigenstate top-quark (which has $SU(2)\times U(1)$ gauge
interactions at site $(1,1)$) \cite{Arkani-Hamed:2001nc,
Arkani-Hamed:2002pa}. Upon diagonalizing the resulting mass matrix,
the expected order-one Yukawa coupling, $y_t$, of the Higgs to the
top-quark is generated so long as the nearest neighbor couplings are
of order one and the lightest vector-fermion mass $m$ satisfies $m/f
\simeq y_t$.

One might imagine that the bounds of Section 2 could be evaded in a
different class of models in which the light fermions are also spread
out in theory space, perhaps with ``families'' of $SU(3)$ vector
fermions.  Even in this case, however, the crucial flavor-violating
couplings are still Yukawa couplings between an ordinary fermion at
the site $(1,1)$ and the appropriate component of a vector fermion at
an adjacent site.  The bounds described in Section
\ref{sec:flavor} apply to these couplings and constrain the
corresponding models.

\section{Flavor and Fine-Tuning in Theory Space}

In order to understand the implications of the lower bound from
flavor physics, we will examine an upper bound imposed by the wish to
avoid fine-tuning in the Higgs boson masses. As noted before, the chiral
symmetries of theory space imply that the leading contributions to the
Higgs boson masses are {\it finite} contributions arising from the
Coleman-Weinberg potential. The rules of power-counting are easily
modified in this case \cite{Arkani-Hamed:2001nc, Arkani-Hamed:2002pa}
in order to estimate the size of these finite contributions to
parameters in the low-energy theory.  In particular, the size of these
contribution is the same as that in a standard scalar Higgs model with
a cutoff equal to the mass of the lowest appropriate resonance. 

For example, gauge boson loop corrections to the Higgs boson masses
are of order \cite{Arkani-Hamed:2001nc, Arkani-Hamed:2002pa}
\beq
\delta m^2_H \simeq {e^2 \over 16 \pi^2\sin^2\theta_W} 
\left({g f \over N}\right)^2
\simeq \left({\alpha\over 4\pi \sin^2\theta_W}\right)^2 \Lambda^2~,
\eeq
where the mass of the first vector resonance (of order $g f/N$) plays
the role of the cutoff of the low-energy theory, and we have assumed
that $g ={\cal O}(N e/\sin\theta_W)$ in order to yield the appropriate
low-energy weak coupling constant.  Here, and in the rest of this
paper, we take $\Lambda \simeq 4\pi f$ which corresponds to $\kappa =
4\pi$ above. The size of these finite corrections to the Higgs boson
mass must be compared to the desired low-energy mass-squared given in
eqn. (\ref{higgsmsq}). To avoid fine-tuning, we require that
\beq
\left|{\delta m^2_H \over m^2_H}\right| \laem 1~,
\eeq
which yields
\beq
\Lambda \laem \left({4 \pi \sin^2\theta_W \over \alpha}\right)\,
{\sqrt{\lambda}\, v\over N} \approx {108\,{\rm TeV}\,\sqrt{\lambda}\over N}~.
\eeq
If gauge-boson loop corrections were the only issue, the cutoff could
be taken to be of order 100 TeV without any fine-tuning.

However, the most important corrections to the Higgs boson masses
arise from the interactions added to give rise to the top-quark
mass. The fermion loop Coleman-Weinberg contribution to the Higgs
mass-squared is of order
\beq
|\delta m^2_H | \simeq {N_c y^2_t m^2 \over 16 \pi^2} \approx 
{N_c y^4_t \over (16 \pi^2)^2}\Lambda^2~,
\label{fermioncw}
\eeq
where $N_c=3$ accounts for color.  In this case, the absence of
fine-tuning ($\delta m^2_H/m^2_H \laem 1$) implies
\beq
\Lambda \laem {16\pi^2 \sqrt{\lambda} v \over \sqrt{N_c} y^2_t N}
\approx {22\,{\rm TeV}\,\sqrt{\lambda}\over N}~.
\label{lambdabound}
\eeq

Comparing eqns. (\ref{lambdabound}) and (\ref{eq:Dboundn}) we see that
for $N=2$ fine-tuning on the order of 1\% is required if the bound
from $\Delta C = 2$ mixing is to be satisfied.  If the bound from CP
violation (\ref{eq:Ebound}) must also be satisfied, the fine-tuning
required is of order .04\% .

\section{Isospin Violation}

A crucial issue in all composite Higgs models is the size of
weak-isospin violation \cite{Chivukula:1996sn,Chivukula:1997iw,
Chivukula:1996rz,Chivukula:1999az}. Recall that the standard
one-doublet Higgs model has an accidental custodial isospin symmetry
\cite{custodial}, which naturally implies that the weak-interaction
$\rho$-parameter is approximately one. While all $SU(2) \times U(1)$
invariant operators made of a single scalar-doublet field that have
dimension less than or equal to four automatically respect custodial
symmetry, terms of higher dimension that arise from the underlying
physics at scale $\Lambda$ in general will not. Furthermore, the
interaction given in eqn. (\ref{twohiggs}) does not respect custodial
symmetry. However, the effect of these interactions is to introduce
custodial violation in the spectrum of Higgs boson masses and
therefore only affects the weak interaction $\rho$ parameter at
one-loop.

The embedding of $SU(2)\times U(1)$ in a global $SU(3)$ interaction is
identical to the symmetry structure of the ``Banks model'', which is
known to give rise to isospin violation \cite{chiggs}. This violation
is most directly understood by expanding the kinetic energy terms in
eqn. (\ref{kinetic}) to fourth-order in the pion fields. Keeping only
the terms involving $\phi_u$ and $\phi_v$, we find the isospin
violating interactions
\beq
{\cal L}_{kin} \supset - {1\over 6 N f^2} 
\left[(\partial_\mu \phi^\dagger_u \phi_u)^2 
- (\partial_\mu \phi^\dagger_u \phi_u)
(\phi^\dagger_u \partial^\mu\phi_u) +
(\phi^\dagger_u \partial^\mu\phi_u)^2\right] + u \leftrightarrow v~.
\eeq
Writing the vevs of the Higgs fields as 
\beq
\langle\phi_u \rangle = \left(
\begin{array}{c}
0 \\
{v\cos\beta \over \sqrt{2}}
\end{array}
\right)
\hspace{2cm}
\langle\phi_v \rangle = \left(
\begin{array}{c}
0 \\
{v\sin\beta \over \sqrt{2}}
\end{array}
\right)~,
\eeq
we find the contribution
\beq
\Delta \rho^\star  = \alpha \Delta T = {v^2 \over 4 N^2 f^2}
\left( 1- {\sin^2 2\beta \over 2}\right)~.
\label{deltarho}
\eeq
Current limits derived from precision electroweak observables
\cite{Chivukula:1999az} require that $\Delta T \laem 0.5$ at 95\%
confidence level for a Higgs mass less than 500 GeV. The
bound in eqn. \ref{deltarho} implies that
\beq
\Lambda \simeq 4\pi f \gaem 
{25\,{\rm TeV} \over N} \left(1-{\sin^2 2\beta \over 2}\right)^{1/2}~.
\label{lambdaisobound}
\eeq
Comparing this with eqn.(\ref{lambdabound}), we see that the underlying
strong dynamics cannot be at energies much less than 10 TeV,
even if the high-energy theory contains approximate flavor and CP
symmetries that nullifies the limits of (\ref{eq:Dbound}) and
(\ref{eq:Ebound}).

\section{Discussion}

In this paper, we have shown that the size of flavor violating
interactions arising generically from underlying flavor dynamics in
composite Higgs models constrain the compositeness scale to be at
least 75 TeV.  This bound applies not only to the original composite
higgs models \cite{chiggs}, but also to the recently developed
``theory space'' models \cite{Arkani-Hamed:2001nc,
Arkani-Hamed:2002pa}.  For theory space models based on an $N\times N$
toroidal lattice, the lower limit is $\Lambda \gaem 75\,{\rm TeV}
\sqrt{N}$, so that the bound is 105 TeV for $N=2$.  On the other hand,
if fine-tuning of the higgs mass is to be avoided in such models,
$\Lambda \laem 22\,{\rm TeV} \sqrt{\lambda}/N$; preventing
flavor-changing neutral currents then leads to fine-tuning at the
level of $10/N^3\,$\%.  We have also seen that the lower limit on
$\Lambda$ derived from considering weak isospin violation are somewhat
weaker than those from FCNC, while those from CP-violation in the
neutral Kaon system are potentially much stronger.

It is also interesting to note how one might construct models that are
not constrained by the bounds discussed in this paper.  In order to
produce the appropriate Yukawa couplings without potentially large
effects in neutral-meson mixing, the underlying flavor or strong
dynamics must incorporate additional structure. First, it may be
possible to construct a theory in which the charm mass-eigenstates are
eigenstates of the corresponding flavor gauge-interactions. In this
case, no $\Delta C=2$ interactions arise at the scale relevant for
producing the charm-quark yukawa couplings. Since Cabibbo mixing
exists, however, such interactions {\it will} necessarily arise at the
scale relevant for strange-quark mass generation, yielding the result
of eqn. (\ref{eq:fcncbound}).  Second, the
underlying strong dynamics could potentially be arranged to have a
different scaling behavior, analogous to ``walking technicolor''
\cite{walking}.  In this case one might have Yukawa couplings of order
$\Lambda/M$ rather than the square of that ratio.  Or third, the
underlying flavor dynamics could incorporate an approximate GIM
symmetry \cite{technigim,ctsm}. Similarly, if the underlying dynamical
theory incorporated an approximate CP symmetry, then the low-energy
theory would not necessarily make the dangerously large contributions
to $\varepsilon$ discussed here.

In summary, we have seen that the low-energy structure of the
composite Higgs model alone is not sufficient to eliminate potential
problems with flavor-changing neutral current or excessive CP
violation; solving those problems requires additional information or
assumptions about the symmetries of the underlying strong dynamics.

{\bf Note Added:} After the completion of this manuscript, two minimal
composite Higgs models have recently been proposed \cite{littlei,
littleii}.  As noted by those authors, the constraints discussed in
this note are relevant to the new models as well.


\centerline{\bf Acknowledgments}

We thank Andrew Cohen, Nima Arkani-Hamed, Hong-Jian He, Ken Lane, 
and Martin Schmaltz for helpful discussions. {\em This work was
supported in part by the Department of Energy under grant
DE-FG02-91ER40676 and by the National Science Foundation under grant
PHY-0074274.}


\bibliography{isotheory}
\bibliographystyle{h-elsevier}

\end{document}